\documentclass[twocolumn,pra,showpacs]{revtex4}
\usepackage{amssymb}
\usepackage{amsmath}
\usepackage{graphicx}

\begin{document}

\title{Coherent control of photon transmission :\\
slowing light in coupled resonator waveguide doped with $\Lambda $ Atoms}
\author{Lan \surname{Zhou}}
\affiliation{Department of Physics, Hunan Normal University, Changsha 410081, China}
\affiliation{Institute of Theoretical Physics, Chinese Academy of Sciences, Beijing,
100080,China}
\author{Jing \surname{Lu}}
\affiliation{Department of Physics, Hunan Normal University, Changsha 410081, China}
\author{C. P. \surname{Sun}}
\email{suncp@itp.ac.cn}
\homepage{http://www.itp.ac.cn/~suncp}
\affiliation{Institute of Theoretical Physics, Chinese Academy of Sciences, Beijing,
100080, China}

\begin{abstract}
In this paper, we propose and study a hybrid mechanism for coherent
transmission of photons in the coupled resonator optical waveguide
(CROW) by incorporating the electromagnetically induced transparency
(EIT) effect into the controllable band gap structure of the CROW.
Here, the configuration setup of system consists of a CROW with
homogeneous couplings and the artificial atoms with $\Lambda$-type
three levels doped in each cavity. The roles of three levels are
completely considered based on a mean field approach where the
collection of three-level atoms collectively behave as two-mode spin
waves. We show that the dynamics of low excitations of atomic
ensemble can be effectively described by an coupling boson model.
The exactly solutions show that the light pulses can be stopped and
stored coherently by adiabatically controlling the classical field.
\end{abstract}

\pacs{42.70.Qs,42.50.Pq,73.20.Mf,  03.67.-a}
\maketitle

\section{Introduction}

Electromagnetically induced transparency (EIT) is a phenomenon that
usually occurs for atomic ensemble as an active mechanism to slow
down or stop laser pulse completely
\cite{EIT97,EIT97-1,EIT01,EIT02}. Usually, the EIT effect happens in
the so-called $\Lambda $-type atomic system, which contains two
lower states with separate couplings to an excited state via two
electromagnetic fields (probe and control light). When the
absorption on both transitions is suppressed due to destructive
interference between excitation pathways to the upper level, the
medium becomes transparent with respect to the probe field.

Most recently, an EIT-like effect has been displayed in the
experiment via all optical on-chip setups with the coupled resonator
optical waveguide (CROW) \cite{Fanprl06,nat441}. The bare CROW for
photons behaves as the tight-binding lattice with band structure for
electrons, thus the CROW forms a new type photonic crystal. It was
discovered that, by coupling each resonator in the CROW to an extra
cavity, the resonate spectral line is shift and the band width is
compressed, thus the propagating of light pulses is stopped and the
information carried by light is stored
\cite{Fanprl04,Fanprl05,Maol29}. The scheme of stopping, storing and
releasing light is also theoretically proposed and analyzed for
quantum-well Bragg structures which form a one-dimensional resonant
photonic bandgap structures \cite{JOSAB1}.

Actually, with the help of modern nano-fabrication technology, the
hybrid structure, i.e., an array of coupled cavities with doping
artificial atoms can be implemented experimentally with a photonic
crystal or other semiconductor systems. By making use of such hybrid
system \cite{Bqp06159,Pqp06097}, Mott insulator and superfluid state
can emerge in different phases of the polaritons formed by dressing
the doping atoms with the gapped light field. Also the hybrid system
of a two-dimensional array of coupled optical cavities in the
photon-blockade regime will undergo a characteristic Mott insulator
(excitations localized on each site) to superfluid (excitations
delocalized across the lattice) quantum phase transition
\cite{lqpt}. Such a coplanar hybrid structure based on
superconducting circuit, has been proposed by us \cite{scp06085} for
the coherent control of microwave - photons propagating in a coupled
superconducting transmission line resonator (CTLR) waveguide
\cite{yale,yale04}.

%%%%%%%%%%%%%%%%%%%%%%%%%%%%%%%%%%%%%%%%%%%%%%%%%%%%%%%%%%%%%%%%%%%%%%%%%%%%
\begin{figure}[tbp]
\includegraphics[bb=78 421 525 701, width=7 cm, clip]{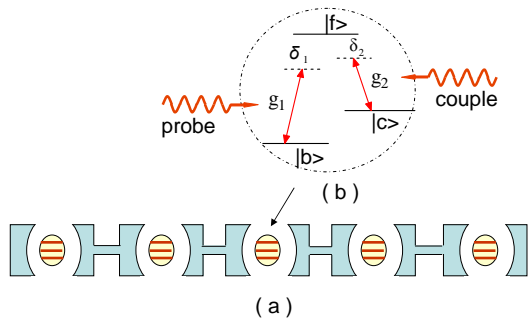}
\caption{\textit{(Color on line)} Illustrative configuration for controlling
light propagation in a coupled resonators waveguide by doping atom systems:
(a) A schematic of the coupled resonator optical waveguide (CROW) doped by
three level systems. The CROW can be realized as a array of periodic defects
in the photonic bandgap materials, or a waveguide with coupled
superconducting transmission line resonators; (b). The doping three level
atoms interacting with quantized probe light and the classical couple light.
The artificial atoms can be realized as the quantum dot, or the charge and
flux qubits.}
\label{fig:1}
\end{figure}
%%%%%%%%%%%%%%%%%%%%%%%%%%%%%%%%%%%%%%%%%%%%%%%%%%%%%%%%%%%%%%%%%%%%%%%%%%%%%

By making use of the two time Green function approach, we studied
the coherent control of photon transmission along the homogeneous
CROW by doping two-level atoms\cite{FMHU}. Usually, to realize the
controllable and robust two-level system, a three-level atom is used
to reduce an effective two level structure through the stimulated
Raman mechanism, which is two photon process decoupling the two
direct transitions to upper energy level in the case with large
detuning. Then the induced coupling between two lower energy levels
can be obtained by the adiabatic elimination of the upper energy
level.

In this paper, we study the photon transmission in a homogeneous
CROW controlled coherently by doped three-level $\Lambda $-type
systems, where the upper energy level is not eliminated
adiabatically. To consider the coherent roles of three energy levels
directly, we use the mean field approach to deal with the collective
excitations of all spatial distributed $\Lambda $ -type atoms as two
independent bosonic modes of quasi-spin-waves \cite{Jin-GR}. These
quasi-spin-waves due to interacting with the cavity modes in CROW
can change the photonic band structure of CROW so that the
dispersion relation exhibits some exotic feature - a slow (and even
zero velocity) light pulses can emerge by some appropriate coherent
control of light-atom couplings.

This paper is organized as follow: In Sec. \ref{sec:one}, we
describe our model - the homogeneous CROW with each cavity doping a
$\Lambda $-type three-level atom. By the mean field approach in
terms of spin wave excitations, in Sec.\ref{sec:one1}, we derive
down the effective Hamiltonian for the hybrid structure. In Sec.
\ref{sec:two}, we diagonalize the effective Hamiltonian to determine
eigenfrequencies of this hybrid photon-atom system. The
quasiparticles - polaritons are introduced to describe the
excitations of this system. Then, in Sec. \ref{sec:three}, we
discuss how the doping atoms modify the band structure of the CROW,
and show how to store the information of incident pulse by adjusting
the intensity of the control radiation in EIT. The absorption and
dispersion of the atomic medium to the slow light pulses are studied
in Sec. \ref{sec:four}. We make our conclusion and give remarks
shortly in sec. \ref{sec:five}.

\section{\label{sec:one} Model setup and motivations}

The hybrid system that we considered is shown in Fig.\ref{fig:1}.
This system consists of $N$ single-mode cavities with homogeneous
nearest-neighbor interaction, which form a one-dimensional array of
cavities. Each single-mode cavity has the same resonance frequencies
$\omega _{0}$. There are three practical systems to implement such
array of cavities \cite{JOSAB2}: 1) a periodic array of coupled
Fabry-Perot cavities; 2) the coupled microdisk or microring
resonators; 3) the coupled defect modes in photonic crystals, where
the bandgap cavities are formed when the periodicity of photonic
crystal is broken periodically \cite{Stefa98,Bayin00}. The
intercavity photon hopping is due to the evanescent coupling
pathways between the cavities. In the coupled Fabry-Perot cavities
and the microdisk or microring resonators, the doped system can be
the natural atom. For photonic crystals, the photonic bandgap
material is fabricated in diamond, the doping systems can be
realized as some ion-implanted NV centers \cite{lqpt}. Another
promising candidate for electromagnetically controlled quantum
device is based on superconducting circuit \cite{scp06085}, where
the CROW is realized by the superconducting waveguide with coupled
transmission line resonators \cite{yale,yale04}, while the doping
systems are implemented by the biased Cooper pair boxes (CPBs) (or
called charge qubits), or the current biased flux qubits.

Generally, we use $a_{j}^{\dag }$ ($a_{j}$) to denote its creation
(annihilation) operator of the $j$th cavity. In each cavity, the two lower
levels $\left\vert b\right\rangle $ and $|c\rangle $ are excited to the
upper level $\left\vert f\right\rangle $ by the quantized field and the
coupling field respectively. The energy level spacing between the upper
level $\left\vert f\right\rangle $ and the ground state $\left\vert
b\right\rangle $ is denoted by $\omega _{fb}=\omega _{f}-\omega _{b}$. This
two-level atomic transition couples to quantized radiation modes of the
waveguide cavities with coupling constant $g_{1}$. The energy difference
between the upper level $\left\vert f\right\rangle $ and the metastable
lower state $\left\vert c\right\rangle $ is denoted by $\omega _{fc}=\omega
_{f}-\omega _{c}$. The atomic transition from $\left\vert f\right\rangle $
to $\left\vert c\right\rangle $ is driven homogeneously by an classical
field of frequency $\Omega $ with coupling constant $g_{2}$.

Let $J$ be the nearest-neighbor evanescent coupling constant of
intercavity. The model Hamiltonian $H=H_{C}+H_{A}+H_{AC}$ consists
of three parts, the cavity part with intercavity photon hoppings,
\begin{equation}
H_{C}=\sum_{j}^{N}\omega _{0}a_{j}^{\dag }a_{j}+J\sum_{j=1}^{N}a_{j}^{\dag
}a_{j+1}+h.c.,
\end{equation}%
the free atom part,
\begin{equation}
H_{A}=\sum_{j}^{N}\left( \omega _{f}\sigma _{ff}^{j}+\omega _{b}\sigma
_{bb}^{j}+\omega _{c}\sigma _{cc}^{j}\right) ,
\end{equation}%
and the localized photon-atom interaction part
\begin{equation}
H_{AC}=\sum_{j}^{N}\left( g_{1}\sigma _{fb}^{j}a_{j}+g_{2}e^{-i\Omega
t}\sigma _{fc}^{j}+h.c.\right)   \label{1}
\end{equation}%
Here, the quasi-spin operators $\sigma _{\alpha \beta }^{j}=|\alpha \rangle
_{j}\langle \beta |$ ($\alpha ,\beta =f,b,c$) for $\alpha \neq \beta $
describe the atomic transitions among the energy levels of $|f\rangle $, $%
|b\rangle $ and $|c\rangle $. In practical experiments, coupling constants $%
g_{i}$ and $J$ depend on the positions of atoms. In this paper, we
take uniform $g$ and $J$ for simplicity. Actually, a small
difference of couplings is unavoidable in the practical
implementation of the present setup, but there should been no
principle difficulty in modern fabrication technique to achieve
quasi uniform coupling \cite{sunprb71}. Theoretically, the small
fluctuations of coupling constants are innocuous and do not change
the results of this paper qualitatively.

%%%%%%%%%%%%%%%%%%%%%%%%%%%%%%%%%%%%%%%%%%%%%%%%%%%%%%%%%%%%%%%%%%%%%%%%%%%%
\begin{figure}[tbp]
\includegraphics[width=8 cm]{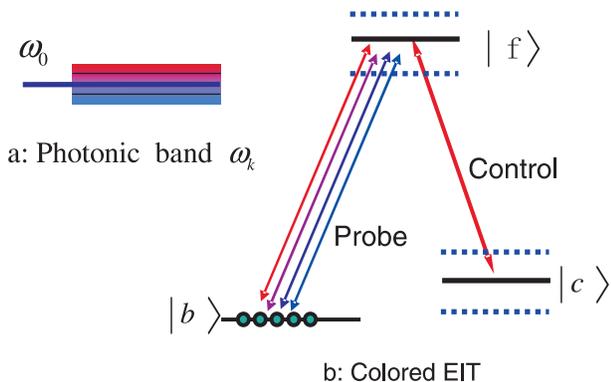}
\caption{\textit{(Color on line)} Electromagnetically induced transparency
(EIT) effect in the coupled resonator optical waveguide (CROW) with photonic
band structure: The problem is equivalent to that a multi-mode optical pulse
with different color component couples between two energy levels
near-resonantly. The strong light split the bare energy levels. The EIT
phenomenon emerges when the band structure can match these split. This
mechanism can well controls light propagation in the CROW by doping a
three-level system. }
\label{fig:2}
\end{figure}
%%%%%%%%%%%%%%%%%%%%%%%%%%%%%%%%%%%%%%%%%%%%%%%%%%%%%%%%%%%%%%%%%%%%%%%%%%%%%

To illustrate our motivation using this complex hybrid structure, we may
recall the fundamental principle for the EIT phenomenon briefly. Usually, a
absorption region occurs to a weak probe light when it passes through a
medium, but in the presence of the control light, a transparency
\textquotedblleft window\textquotedblright\ appears in the probe absorption
spectrum. Here the probe light is not of single color since the photon
propagating in the CROW has a photonic band. To consider whether or not EIT
phenomenon emerges in this band-gap structure, we should match the photonic
band structure with the splits of the energy level spacing between $%
|f\rangle $ and $|c\rangle $ (see the Fig.2).

As for this hybrid structure with EIT effect, it is well known that, among
varieties of theoretical treatments of EIT, an approach for EIT is the
\textquotedblleft dressed state\textquotedblright\ picture, wherein the
Hamiltonian of the system plus the light field is diagonalized firstly to
give rise to a Autler-Townes like splitting \cite{Autler} in the strong
coupling limit with the control field. Then the Fano like interference \cite%
{Fano124} between the dressed states results in EIT. Between the
doublet peaks of the absorption line, a transparency window emerges
as the quantum probability amplitudes for transitions to the two
lower states interference. In the CROW, the emitted and absorbed
photons can also be constrained by the photonic band structure.
Here, the single and two photon resonances in EIT for a given
Autler-Townes like splitting should be re-considered to match the
band structure of the CROW. Particularly, we need to generalize the
polariton approach to describe the stopped and stored light schemes.
Here, the photons of the probe beam only within the photonic band
can be coherently ``transformed'' into ``dark state polaritons'' ,
which are the dressed excitations of atom ensemble.

\section{\label{sec:one1}Collective Excitations with Electromagnetically
Induced Transparency Effect}

In order to study the novel EIT effect in the CROW, we use the mean field
approach that we developed for the collective excitation of an atomic
ensemble with a ordered initial state \cite{sunprl91}. This approach for EIT
can be understood as a fully-quantum theory, which not only gives these
results about slow light propagation that can be given by semi-classical
approach, but also emphasizes the quantum states of photon and the atomic
collective excitations - quasi-spin waves, which are crucial for quantum
information processing, such as quantum memory or storage

Let $\ell $ be the distance between the nearest-neighbor cavities. The
Fourier transformation
\begin{eqnarray}
A_{k} &=&\frac{1}{\sqrt{N}}\sum_{j}\sigma _{bf}^{j}\exp (ik\ell j) \\
C_{k} &=&\frac{1}{\sqrt{N}}\sum_{j}\sigma _{bc}^{j}\exp (ik\ell j)
\end{eqnarray}%
and its conjugate $A_{k}=(A_{k}^{\dag })^{\dag }$, $C_{k}=(C_{k}^{\dag
})^{\dag }$ \ define the boson-like operators to describe the collective
excitation from $|b\rangle $ to $|f\rangle $ and from $|b\rangle $ to $%
|c\rangle $ respectively. In the large $N$ limit, and under the low
excitation condition that there are only a few atoms occupying $|f\rangle $
or $|c\rangle $, the quasi-spin-wave excitations behave as bosons since they
satisfy the bosonic commutation relations
\begin{eqnarray}
\lbrack A_{k},A_{k}^{\dag }] &=&1,[C_{k},C_{k}^{\dag }]=1.  \notag \\
\lbrack A_{k},C_{k}] &=&0, \\
\lbrack A_{k},C_{k}^{\dag }] &=&-\frac{T_{-}}{N}\rightarrow 0  \notag
\end{eqnarray}%
Thus these quasi-spin-wave low excitations are independent of each other.
Here, the collective operators
\begin{eqnarray}
T_{-} &=&\sum_{j}^{N}\sigma _{cf}^{j},T_{+}=\left( T_{-}\right) ^{\dag } \\
T_{3} &=&\frac{1}{2}\sum_{j}\left( \sigma _{ff}^{j}-\sigma _{cc}^{j}\right)
\end{eqnarray}%
generates the $SU(2)$ algebra.

In a rotating frame with respect to the 0'th order Hamiltonian
\begin{equation*}
H_{0}=\sum_{j}^{N}(\omega _{0}a_{j}^{\dag }a_{j}+\omega _{f}^{\prime }\sigma
_{ff}^{j}+\omega _{b}\sigma _{bb}^{j}+\omega _{c}\sigma _{cc}^{j})
\end{equation*}%
we achieve the coupling boson mode with the model Hamiltonian $%
H=\sum_{k}H_{k}:$
\begin{eqnarray}
H_{k} &=&\delta _{2}A_{k}^{\dag }A_{k}+\Omega _{k}a_{k}^{\dag }a_{k}  \notag
\\
&&+g_{1}A_{k}^{\dag }a_{k}+g_{2}A_{k}^{\dag }C_{k}+h.c.  \label{4}
\end{eqnarray}%
where we have used the Fourier transformation
\begin{equation*}
\hat{a}_{k}=\frac{1}{\sqrt{N}}\sum_{j}\hat{a}_{j}\exp (ik\ell j).
\end{equation*}%
Here, $\omega _{f}^{\prime }=\omega _{f}-\delta _{1}$, $\delta
_{1}=\omega _{fb}-\omega _{0}$ is detuning between the quantized
mode and the transition frequency $\omega _{fb}$, and $\delta
_{2}=\omega _{fc}-\Omega $ is detuning between the classical field
and the transition frequency $\omega _{fc}$. The original band
structure is characterized by the dispersion relation
\begin{equation}
\Omega _{k}=\delta _{2}-\delta _{1}+2J\cos \left( k\ell \right) .  \label{5}
\end{equation}%
Obviously the photonic band is centered at $k=\pi /(2\ell )$.

To enhance the coupling strength between the probe field and atoms, we can
dope more, say $N_{A}$,  identical noninteractive three-level $\Lambda $%
-type atoms in each cavity. In this case, the system Hamiltonian is changed
into
\begin{equation*}
H=H_{C}+\sum_{j}(H_{A}^{j}+H_{CA}^{j})
\end{equation*}%
with
\begin{eqnarray}
&&H_{A}^{j}=\omega _{f}s_{ff}^{j}+\omega _{b}s_{bb}^{j}+\omega _{c}s_{cc}^{j}%
\text{,} \\
&&H_{CA}^{j}=g_{1}s_{fb}^{j}a_{j}+g_{2}e^{-i\Omega t}s_{fc}^{j}+\text{H.c.,}
\label{two-1}
\end{eqnarray}%
where, in each cavity,
\begin{equation}
s_{\alpha \beta }^{j}=\sum_{l}\sigma _{\alpha \beta j}^{l}
\end{equation}%
denote the collective dipole between $|\alpha \rangle $ and $|\beta \rangle $
for $\alpha \neq \beta $.

For each cavity, the collective effect of doped three-level atoms can be
described by quasi-spin-wave boson operators
\begin{equation}
A_{j}=\frac{s_{bf}^{j}}{\sqrt{N_{A}}},C_{j}=\frac{s_{bc}^{j}}{\sqrt{N_{A}}},
\label{two-2}
\end{equation}%
which create two collective states $|1_{c}\rangle _{j}=C_{j}^{\dag }|\nu
\rangle $ and $|1_{f}\rangle _{j}=A_{j}^{\dag }|\nu \rangle $ with one
quasi-particle excitations. Here $|\nu \rangle =|b_{1},b_{2},\ldots
,b_{N_{A}}\rangle $ is the collective ground state with all $N_{A}$ atoms
staying in the ground state $|b\rangle $. In low excitation and large $N_{A}$
limit, the two quasi-spin-wave excitations behave as two bosons\cite%
{sunprl91}, and they satisfy the bosonic commutation relations%
\begin{equation}
\lbrack A_{j},A_{j}^{\dag }]=1,[C_{j},C_{j}^{\dag }]=1,
\end{equation}
and $[A_{j},C_{j}]=0$. The commutation relations between $A_{j}$ and $C_{j}$
means that, in each cavity, the two quasi-spin-wave generated by $N_{A}$
three-level $\Lambda $-type atoms are independent of each other.

In the interaction picture with respect to
\begin{equation*}
H_{0}=\omega _{0}\sum_{j}^{N}a_{j}^{\dag }a_{j}+\sum_{j}^{N}\left[ \omega
_{f}^{\prime }s_{ff}^{j}+\omega _{b}s_{bb}^{j}+\omega _{c}s_{cc}^{j}\right] ,
\end{equation*}
and by the Fourier transformations
\begin{equation}
F_{k}=\sum_{j}\frac{F_{j}}{\sqrt{N_{A}}}e^{ik\ell j}  \label{two-3}
\end{equation}
for $F=a,A$ and $C$ et al, the interaction Hamiltonian reads as $%
V=\sum_{k}V_{k}$:
\begin{eqnarray}
V_{k} &=&\epsilon _{k}a_{k}^{\dag }a_{k}+\delta _{2}A_{k}^{\dag }A_{k}
\notag \\
&&+G_{1}A_{k}^{\dag }a_{k}+g_{2}A_{k}^{\dag }C_{k}+h.c.  \label{two-4}
\end{eqnarray}
where
\begin{equation}
\epsilon _{k}=2J\cos \left( k\ell \right) +\delta _{2}-\delta _{1}
\label{eq:epsi}
\end{equation}
is the dispersion relation of CROW. Here, the effective photonic band-spin
wave coupling $G_{1}=g_{1}\sqrt{N_{A}}$ is $\sqrt{N_{A}}$ times enhancement
of $g_{1}$ and thus result in a strong coupling.

We also notice that the $SU(2)$ algebra defined by the quasi-spin operators $%
T_{-},T_{+}$ and $T_{3}$ in the coordinate space can also be realized in the
momentum space through the Fourier transformations as
\begin{eqnarray}
(T_{-})_{k} &=&A_{k}^{\dag }C_{k},(T_{+})_{k}=C_{k}^{\dag }A_{k}  \notag \\
(T_{3})_{k} &=&\frac{1}{2}(C_{k}^{\dag }C_{k}-A_{k}^{\dag }A_{k})
\end{eqnarray}%
This means the interaction Hamiltonian possesses a intrinsic dynamic
symmetry described by a large algebra containing $SU(2)$ as subalgebra.
Technologically this observation will help us to diagonalize the Hamiltonian
Eq.(\ref{two-4}) as follows.

\section{\label{sec:two} Dressed collective states: polaritons}

In each cavity the strong couplings will coherently mix the photon and the
artificial atoms to form dressed states. The collective effect of these
dressed states can make the collective excitations, which behave as bosons
(called polaritons) \ in the low excitation limit. Mathematically we write
the boson operator $a_{k}$, $A_{k}$ and $C_{k}$ as an operator-valued vector
\begin{equation*}
\vec{b}_{k}=\left(
\begin{array}{c}
a_{k} \\
A_{k} \\
C_{k}%
\end{array}%
\right) .
\end{equation*}%
In terms of those operator-valued vectors \{$\vec{b}_{k}$\}, the interaction
Hamiltonian $V_{k}$ can be re-written as%
\begin{equation*}
V_{k}=\vec{b}_{k}^{\dag }M\vec{b}_{k},
\end{equation*}%
where
\begin{equation*}
M=\left[
\begin{array}{ccc}
\epsilon _{k} & G_{1} & 0 \\
G_{1} & \delta _{2} & g_{2} \\
0 & g_{2} & 0%
\end{array}%
\right] .
\end{equation*}

Now we solve eigenvalue problem of the matrix $M$. Then $V_{k}$ can be
diagonalized to construct the polariton operators, which is described by the
linear combination of the quantized electromagnetic field operators and
atomic collective excitation operator of quasi-spin waves. The three real
eigenvalues of $M$
\begin{eqnarray}
\lambda _{k}^{[1]} &=&\beta _{+}^{[k]}+\beta _{-}^{[k]}+\frac{1}{3}(\epsilon
_{k}+\delta _{2})  \notag \\
\lambda _{k}^{[2]} &=&\varkappa \beta _{+}^{[k]}+\varkappa ^{2}\beta
_{-}^{[k]}+\frac{1}{3}(\epsilon _{k}+\delta _{2}) \\
\lambda _{k}^{[3]} &=&\varkappa ^{2}\beta _{+}^{[k]}+\varkappa \beta
_{-}^{[k]}+\frac{1}{3}(\epsilon _{k}+\delta _{2})  \notag
\end{eqnarray}%
are written in terms of $\varkappa =(-1+i\sqrt{3})/2$ and
\begin{equation*}
\beta _{\pm }^{[k]}=\sqrt[3]{-\frac{q}{2}\pm \sqrt{\left( \frac{q}{2}\right)
^{2}+\left( \frac{p}{3}\right) ^{3}},}
\end{equation*}%
\begin{eqnarray*}
p &=&-\frac{1}{3}\epsilon _{k}^{2}+\frac{1}{3}\delta _{2}\epsilon
_{k}-G_{1}^{2}-g_{2}^{2}-\frac{1}{3}\delta _{2}^{2}, \\
q &=&\frac{1}{27}\left[ 3\delta _{2}\epsilon _{k}^{2}-2\epsilon
_{k}^{3}+\left( 18g_{2}^{2}-9G_{1}^{2}+3\delta _{2}^{2}\right) \epsilon
_{k}\right. \\
&&\left. -2\delta _{2}^{3}-9G_{1}^{2}\delta _{2}-9g_{2}^{2}\delta _{2}\right]
,
\end{eqnarray*}%
For a nonzero eigenvalue $\lambda _{k}^{[i]}$, the polariton operators can
be defined as
\begin{equation}
P_{k}^{[i]}=\frac{1}{r_{i}}\left[ \frac{G_{1}}{\lambda _{k}^{[i]}-\epsilon
_{k}}a_{k}+A_{k}+\frac{g_{2}}{\lambda _{k}^{[i]}}C_{k}\right] ,
\label{eq:polgp}
\end{equation}%
where
\begin{equation}
r_{i}=\sqrt{\frac{|G_{1}|^{2}}{|\lambda _{k}^{[i]}-\epsilon _{k}|^{2}}+1+%
\frac{|g_{2}|^{2}}{|\lambda _{k}^{[i]}|^{2}}}.
\end{equation}

When the detunings approximately satisfy the resonance transition condition
so that $\epsilon _{k}=0$ for some $k,$ the dark-state polariton can be
constructed as an eigenstate with vanishing eigenvalue. For concreteness, we
first consider the case with the detuning $\delta _{2}=\delta _{1}=0$, which
means that the probe light and the classical field are resonant with the $%
\Lambda $-type atoms in each cavity. The polariton operators at the band
center $k=k_{0}=\pi /(2\ell )$ can be constructed as
\begin{subequations}
\label{eq:darkp}
\begin{eqnarray}
P_{k_{0}}^{[1]} &=&\frac{1}{\sqrt{2}}(A_{k_{0}}-B_{k_{0}})  \label{eq:pop1}
\\
P_{k_{0}}^{[2]} &=&a_{k_{0}}\cos \theta -C_{k_{0}}\sin \theta
\label{eq:pop2} \\
P_{k_{0}}^{[3]} &=&\frac{1}{\sqrt{2}}(A_{k_{0}}+B_{k_{0}})  \label{eq:pop3}
\end{eqnarray}%
with $\tan \theta =G_{1}/g_{2}$, and
\end{subequations}
\begin{equation}
B_{k_{0}}=a_{k_{0}}\cos \theta +C_{k_{0}}\sin \theta .
\end{equation}%
Here, $P_{k_{0}}^{[2]}$ is the dark-state polariton (DSP), which traps the
electromagnetic radiation from the excited state due to quantum interference
cancelling; $B_{k_{0}}$ is called the bright-state polariton\cite{sunprl91}.

For another case, we assume that, in each cavity, the frequency of the probe
light $\omega _{0}$ has a nonzero detuning from the transition frequency $%
\omega _{ab}$, i.e. $\delta _{1}=\Delta \neq 0$. By adjusting the frequency
of the classical field, $\delta _{2}=\Delta $ can be realized, and then the
condition $\epsilon _{k}=0$ is satisfied at the band center. So the
dark-state polariton exists. With the polariton operators
\begin{subequations}
\label{eq:darkQ}
\begin{eqnarray}
Q_{k_{0}}^{[1]} &=&\xi \left[ G_{1}a_{k_{0}}+\frac{\Delta -\alpha }{2}%
A_{k_{0}}+g_{2}C_{k_{0}}\right] ,  \label{eq:poq1} \\
Q_{k_{0}}^{[2]} &=&a_{k_{0}}\cos \theta -C_{k_{0}}\sin \theta ,
\label{eq:poq2} \\
Q_{k_{0}}^{[3]} &=&\xi \left[ G_{1}a_{k_{0}}+\frac{\Delta +\alpha }{2}%
A_{k_{0}}+g_{2}C_{k_{0}}\right] ,  \label{eq:poq3}
\end{eqnarray}%
for
\end{subequations}
\begin{equation*}
\xi =\sqrt{2/\left( \alpha -\Delta \right) \alpha },
\end{equation*}%
the interaction Hamiltonian $V_{k_{0}}$ is diagonalized. Here,
\begin{equation}
\alpha =\sqrt{\Delta ^{2}+4G_{1}^{2}+4g_{2}^{2}}.
\end{equation}%
the DSP $Q_{k_{0}}^{[2]}$ is the specific light-matter dressed states, which
particularly appears in EIT.

Actually, for a probe light with nonzero detuning $\delta _{1}$ and small
band around $k=k_{1}$ ($k_{1}\neq k_{0}$), by adjusting the detuning $\delta
_{2}$ to satisfy $\epsilon _{k_{1}}=0$, at the model $k=k_{1}$, we can also
construct the polariton operators similar to those of Eq. (\ref{eq:darkQ})
with $\delta _{2}$ replacing $\Delta $ and $k_{1}$ replacing $k_{0}$.

\section{\protect\bigskip \label{sec:three}Band structure of polaritons}

\begin{figure}[tbp]
\includegraphics[width=8 cm]{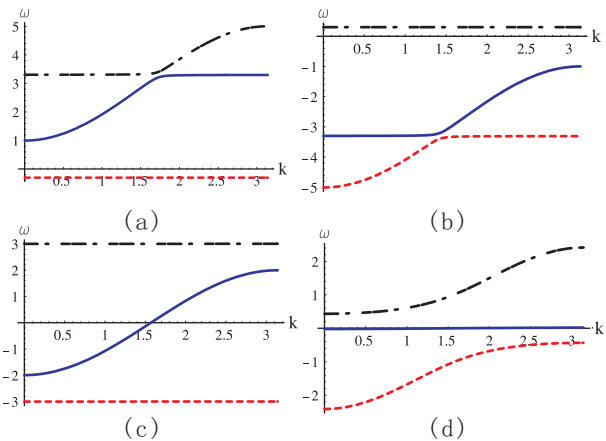}
\caption{\textit{(Color on line)} The band structure at the first excitation
space. Here the eigenfrequency is plotted as a function of the wave vector $k
$. We have set $J=-1$. The other parameter are set as follow, $\protect%
\delta _{1}=0$, (a)$\protect\delta _{2}=3|J|$, $G_{1}=0.1J$, $g_{2}=1.0|J|$,
( b)$\protect\delta _{2}=-3|J|$, $G_{1}=0.1J$, $g_{2}=1.0|J|$. (c)$\protect%
\delta _{2}=0$, $G_{1}=0.1|J|$, $g_{2}=3|J|$, (d)$\protect\delta _{2}=0$, $%
G_{1}=|J|$, $g_{2}=0.1|J|$. The wave vector $k$ is in unit of $1/\ell $.}
\label{fig:band}
\end{figure}

From the above discussion, it can be observed that the spectra of the hybrid
system consists of three bands, and there exists gaps among these three
bands for a non-vanishing $G_{1}$ and $g_{2}$. Since the number of total
excitations
\begin{equation}
N_{k}=a_{k}^{\dag }a_{k}+A_{k}^{\dag }A_{k}+C_{k}^{\dag }C_{k}
\end{equation}%
commutes with $V_{k}$, the number of excitation $N_{k}$ is conserved , while
the numbers $a_{k}^{\dag }a_{k}$, $A_{k}^{\dag }A_{k}$ and $C_{k}^{\dag
}C_{k}$ of different type excitations are mutually convertible by adjusting
some parameters. In Fig.\ref{fig:band} we plot the eigenfrequencies as a
function of the wave vector $k$ in the one excitation subspace.
%%%%%%%%%%%%%%%%%%%%%%%%%%%%%%%%%%%%%%%%%%%%%%%%%%%%%%%%%%%%%%%%%%%%%%%%%%
%%%%%%%%%%%%%%%%%%%%%%%%%%%%%%%%%%%%%%%%%%%%%%%%%%%%%%%%%%%%%%%%%%%%%%%%%%%%%
It can be seen from Figs. \ref{fig:band}(a) and (b) that the bandwidth can
be tuned by adjusting the detuning $\delta _{2}$ and the coupling strength $%
g_{2}$. For a fixed coupling strength $g_{2}$, when $\delta _{2}\ll -|g_{2}|$%
, the lowest band (the red dash line) has a large bandwidth, which ensures
to accommodate the bandwidth of entire pulse; when $\delta _{2}\gg |g_{2}|$,
the bandwidth of the lowest band $W_{0}\approx 0$. Hence for a light pulse
that is a superposition of many $k$-states, its distribution in the $k$
-space can be entirely contained in the photonic band of the CROW by setting
$\delta _{2}\ll -|g_{2}|$. By adiabatically tuning the detuning from $\delta
_{2}\ll -|g_{2}|$ to $\delta _{2}\gg |g_{2}|$, the light pulse can be
stopped. Such kind approach to stoping light with all-optical process has
been investigated theoretically by numerical simulations \cite%
{Fanprl04,Fanprl05,Maol29} and a similar all-optical scheme has already been
realized in a recent experiment \cite{Fanprl06}.

When the light pulse enters the medium, photons and atoms combine to form
excitations known as polaritons. Because the spin wave propagates together
with the light pulse inside the medium, the group velocity of light pulse is
reduced by a large order of magnitude. Thus by analyzing the contribution of
photons to the polaritons, it can be well understood that how the group
velocity of probe field is stopped and revived. For the sake of simplify,
firstly, we focus on the polaritons at the band center and consider the
situation with the resonance transition. The operators of polaritons are the
linear combination of that of photons and atoms with the following form
\begin{subequations}
\label{eq:apdark}
\begin{eqnarray}
P_{k_{0}}^{[1]} &=&\frac{1}{\sqrt{2}}(A_{k_{0}}-a_{k_{0}}\cos \theta
-C_{k_{0}}\sin \theta )  \label{eq:pc1} \\
P_{k_{0}}^{[2]} &=&a_{k_{0}}\cos \theta -C_{k_{0}}\sin \theta
\label{eq:pc2} \\
P_{k_{0}}^{[3]} &=&\frac{1}{\sqrt{2}}(A_{k_{0}}+a_{k_{0}}\cos \theta
+C_{k_{0}}\sin \theta )  \label{eq:pc3}
\end{eqnarray}%
where
\end{subequations}
\begin{eqnarray}
\cos \theta  &=&\frac{g_{2}}{\sqrt{g_{2}^{2}+G_{1}^{2}}} \\
\sin \theta  &=&\frac{G_{1}}{\sqrt{g_{2}^{2}+G_{1}^{2}}}.
\end{eqnarray}

The contribution of photons to dark polaritons can be explicitly analyzed.
It can be obtained that the dark polariton appears like photons with
probability approximately to one when $g_{2}\gg G_{1}$, that is, $%
P_{k_{0}}^{[2]}\approx a_{k_{0}}$. Thus if we initial set $g_{2}\gg G_{1}$,
this means the middle band can accommodate many component of the input
pulse. It is easy to find that when $g_{2}\ll G_{1}$, the contribution of
photons in the polariton becomes purely atomic, that is, $%
P_{k_{0}}^{[2]}\approx C_{k_{0}}$. Thus when the pulse is completely in the
system, the adiabatical performance changes the dark polariton from photons
to atoms and reverse. The similar situation can be found at the second band
under the two photon resonance from Eq. (\ref{eq:darkQ}).

\begin{figure}[tbp]
\includegraphics[width=8 cm]{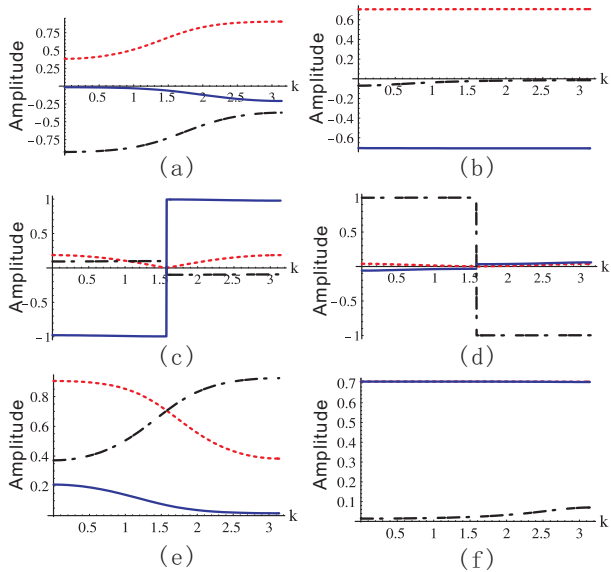}
\caption{\textit{(Color on line)} the contribution of photons and two spin
waves in the polaritons $d_{jk}$ as functions of $k$ for the first
eigenfrequency $\protect\lambda _{k}^{[1]}$ (a) and (b), the second
eigenfrequency $\protect\lambda _{k}^{[2]}$ (c) and (d), the third
eigenfrequency $\protect\lambda _{k}^{[3]}$ (e) and (f). $\protect\delta %
_{1}=\protect\delta _{2}=0$, $J=-1$ . For (a), (c) and (e), we set $G_{1}=|J|
$ and $g_{2}=0.1|J|$; For (b), (d) and (f), we set $G_{1}=0.1|J|$ and $%
g_{2}=3|J|$. The wave vector $k$ is in unit of $1/\ell $}
\label{fig:amp}
\end{figure}

In order to give a general argument, we plot the coefficients before $a_{k}$%
, $A_{k}$ and $C_{k}$ in the polaritons as functions of the momentum index $k
$ respectively in Fig.\ref{fig:amp}. For the convenience of expression, we
denote $d_{jk}^{[i]}$ ($j=1,2,3$) as the coefficients before the operators $%
a_{k}$, $A_{k}$ and $C_{k}$ for different eigenvalues $i=1,2,3$
respectively. From Eq. (\ref{eq:polgp}), the expression of $d_{jk}^{[i]}$
can be obtained
\begin{eqnarray}
d_{1k}^{[i]} &=&\frac{1}{r_{i}}\frac{G_{1}}{\lambda _{k}^{[i]}-\epsilon _{k}}%
,  \notag \\
d_{2k}^{[i]} &=&\frac{1}{r_{i}}, \\
d_{3k}^{[i]} &=&\frac{1}{r_{i}}\frac{g_{2}}{\lambda _{k}^{[i]}}.  \notag
\end{eqnarray}%
In each figure,
%%%%%%%%%%%%%%%%%%%%%%%%%%%%%%%%%%%%%%%%%%%%%%%%%%%%%%%%%%%%%%%%%%%%%%%
%%%%%%%%%%%%%%%%%%%%%%%%%%%%%%%%%%%%%%%%%%%%%%%%%%%%%%%%%%%%%%%%%%%%%%%%
the red dash line represents the magnitude $d_{2k}^{[i]}$ of spin waves
generated by the atomic transition between $|b\rangle $ and $|f\rangle $;
the blue solid line denotes the amplitude $d_{3k}^{[i]}$ of spin waves
between $|b\rangle $ and $|c\rangle $; the black dash-dot line describes the
magnitude $d_{1k}^{[i]}$ of photonic component. It can be observed that for
the incident pulse with momentum distribution around $k=\pi /(2\ell )$,
photons make a large contribution to the polaritons at $G_{1}\gg g_{2}$ than
at the condition $G_{1}\ll g_{2}$. The contribution of photons in the
polariton of the second band, shown in Fig. \ref{fig:amp}(c) and (d), is
modified completely by the coupling strength of light and matter: spin waves
$C_{k}$ take large proportions when $G_{1}\gg g_{2}$ and photons $a_{k}$
have large contributions when $G_{1}\ll g_{2}$. Hence the second band can be
used to convert the quantum information originally carried by photons into
long-lived spin states of atoms.

The characteristic of our hybrid system is that the \textquotedblleft dark
state\textquotedblright can be realized in a straightforward way. This gives
rise to quasi-particles - the dark polariton, which reflects the crucial
idea of the EIT - the coherent population trapping for the quantized probe
field. Actually, a DSP is an atomic collective excitation (quasi-spin wave)
dressed by the quantized probe light. This point can be seen directly from
Eq.(\ref{eq:poq2}). The contributions of light or atoms in DSP can be varied
by adapting the amplitude of the classical field, which has been discussed
in the last paragraph. Thus, in our hybrid system, the DSP offers a possible
control scheme for slowing light. This accessible scheme can be observed
from the change of bandwidth. In Fig. \ref{fig:band}(c) and (d), we plot the
eigenfrequency as a function of the wave vector $k$ in the first excitation
space for a given $\delta _{2}$. It shows that, when $g_{2}\gg G_{1}$, the
bandwidth of the middle band (the blue solid line) has a large bandwidth;
when $g_{2}\ll G_{1}$, the bandwidth of the middle are approximately to
zero; The couplings also make the center of bands away from $\omega _{0}$, $%
\omega _{ab}$ and $\omega _{cb}$ respectively. This fact means that by
tuning the coupling strength from $g_{2}\gg G_{1}$ to $g_{2}\ll G_{1}$
adiabatically, we can stop the input light pulse and then re-emit it. Thus
via selecting a classical field with a suitable frequency, the quantum state
of an input pulse can be converted into the doped three-level atoms simply
by switching off the driving field, and then by turning on the driving
field, the stored information can be retrieved.

To give a concrete example, we consider the resonant transition with $\delta
_{1}=\delta _{2}=0$. In this case, the corresponding group velocities at
each band center are
\begin{eqnarray}
v_{g}^{1}\left[ k_{0}\right]  &=&\frac{G_{1}^{2}}{G_{1}^{2}+g_{2}^{2}}J\ell ,
\\
v_{g}^{2}\left[ k_{0}\right]  &=&\frac{2g_{2}^{2}}{G_{1}^{2}+g_{2}^{2}}J\ell
, \\
v_{g}^{3}\left[ k_{0}\right]  &=&\frac{G_{1}^{2}}{G_{1}^{2}+g_{2}^{2}}J\ell ,
\end{eqnarray}
It can be seen that, at the band center, when $g_{2}\gg G_{1}$, the lowest
band (the red dash line in Fig.\ref{fig:band}(c) and (d)) and the highest
band (the black dash-dot line in Fig. \ref{fig:band}(c) and (d)) exhibit
zero group velocity and zero bandwidth, but the middle band (the blue solid
one in Fig.\ref{fig:band}(c) and (d)) exhibits a large group velocity and a
large bandwidth; in reverse, when $g_{2}\ll G_{1}$, the middle band exhibits
zero group velocity and vanishing bandwidth, but the lowest band and the
highest band exhibit large group velocities and large bandwidths. Hence in
this system, focusing on the middle band, a light pulse can be stopped by
the following process: Initially setting $g_{2}\gg G_{1}$, the middle band
accommodates the entire pulse. After the pulse is completely in this system,
we vary the coupling strength until $g_{2}\ll G_{1}$ adiabatically. The
lowest band also can be used to stop light by tuning $g_{2}$ from $g_{2}\ll
G_{1}$ to $g_{2}\gg G_{1}$.

Finally, we give some estimation about the group velocity according
to the realistic parameters, which are taken for the array of
coupled toroidal microcavities\cite{Pqp06097,Spipra05}. The distance
$\ell$ between the microcavities is $15.69\mu m$ \cite{Fanprl06},
and the evanescent coupling between the cavities $J=1.1\times
10^{7}s^{-1}$. Within each cavity, the coupling strength between the
atom and the quantum field $ g_{1}=2.5\times 10^{9}s^{-1}$, Rabi
frequency $g_{2}=7.9\times 10^{10}s^{-1}$. When $N=10,000$ atoms are
contained in each cavity, the group velocity of light at the second
band center is estimated $31m/s$.

\section{\label{sec:four} Susceptibility analysis for light propagation in
the doped CROW}

When a light beam incidents on an optically active medium, the medium will
give a response to the control light. Usually, the index of refraction can
reach high values near a transition resonance, but the high dispersion
always accompanies with a high absorption in the resonance point. In EIT,
the resonant transition or the two photon resonance renders a medium
transparent over a narrow spectral range within the absorption line. Also in
this transparent window, the rapidly varying dispersion is created, which
leads to very slow group velocity and zero group-velocity. In this section,
we will investigate the dispersion and the absorption property of the gapped
light in our hybrid system. We use the dynamic algebraic method developed
for the atomic ensemble based quantum memory with EIT \cite%
{sunprl91,sunpra69}.

We begin with the Hamiltonian (\ref{two-4}) in the $k$-space representation.
When the atomic decay is considered, we write down the Heisenberg equations
of operators $a_{k}$, $A_{k}$ and $C_{k}$ for each mode $k$
\begin{eqnarray}
\partial _{t}a_{k} &=&-(\gamma +i\epsilon _{k})a_{k}-iG_{1}A_{k}, \\
\partial _{t}A_{k} &=&-\left( \gamma _{A}+i\delta _{2}\right)
A_{k}-iG_{1}a_{k}-ig_{2}C_{k}, \\
\partial _{t}C_{k} &=&-\gamma _{C}C_{k}-ig_{2}A_{k}
\end{eqnarray}%
where we have phenomenologically introduced the damping rate of cavity $%
\gamma $, and the decay rate $\gamma _{A}$, $\gamma _{C}$ of the energy
levels $|f\rangle $ and $|c\rangle $ of the three-level system respectively.
We also assume that
\begin{equation*}
\gamma _{A}\gg \gamma _{C}\gg \gamma .
\end{equation*}

To find the steady-state solution for the above motion equations, it is
convenient to remove the fast varying part of the light field and the atomic
collective excitations by making a transformation
\begin{equation}
F_{k}=\tilde{F}_{k}e^{-i\epsilon _{k}t}
\end{equation}%
for $F_{k}=a_{k}$, $A_{k}$ and $C_{k}$. For the convenience of notation, we
drop the tilde, and then the above Heisenberg equations become
\begin{eqnarray}
\partial _{t}C_{k} &=&\left( i\epsilon _{k}-\gamma _{C}\right)
C_{k}-ig_{2}A_{k},  \label{eq} \\
\partial _{t}A_{k} &=&\left[ i\left( \omega _{k}-\delta _{1}\right) -\gamma
_{A}\right] A_{k}-iG_{1}a_{k}-ig_{2}C_{k},  \notag
\end{eqnarray}%
where $\omega _{k}=2J\cos \left( k\ell \right) $.

The electric field of the quantized probe light with $k$-space
representation
\begin{equation}
E_{k}(t)=\sqrt{\frac{\omega _{0}}{2V\varepsilon _{0}}}a_{k}e^{-i\epsilon
_{k}t}+h.c.
\end{equation}%
results in a linear response of medium, which is described by the
polarization%
\begin{equation*}
\left\langle P_{k}\right\rangle =\left\langle p_{k}\right\rangle \exp
(-i\epsilon _{k})+h.c.
\end{equation*}%
Here,
\begin{equation}
\left\langle p_{k}\right\rangle =\frac{\mu }{V}\sqrt{N_{A}}\left\langle
A_{k}\right\rangle
\end{equation}%
is a slowly varying complex polarization determined by the population
distribution on $|f\rangle $ and $|c\rangle $; $\mu $ denotes the dipole
moment between $|f\rangle $ and $|c\rangle $, and $V$ is the effective mode
volume\cite{bqoptics}. It is also related to the susceptibility $\chi _{k}$
of the $k$-space by
\begin{equation}
\left\langle p_{k}\right\rangle =\epsilon _{0}\chi _{k}\sqrt{\frac{\omega
_{0}}{2V\varepsilon _{0}}}\left\langle a_{k}\right\rangle .
\end{equation}%
The real part $\chi _{k}^{r}$ and imaginary part $\chi _{k}^{i}$ of the
susceptibility correspond to the dispersion and absorption respectively.

%%%%%%%%%%%%%%%%%%%%%%%%%%%%%%%%%%%%%%%%%%%%%%%%%%%%%%%%%%%%%%%%%%%%%%%%%%%%
\begin{figure}[tbp]
\includegraphics[width=8 cm]{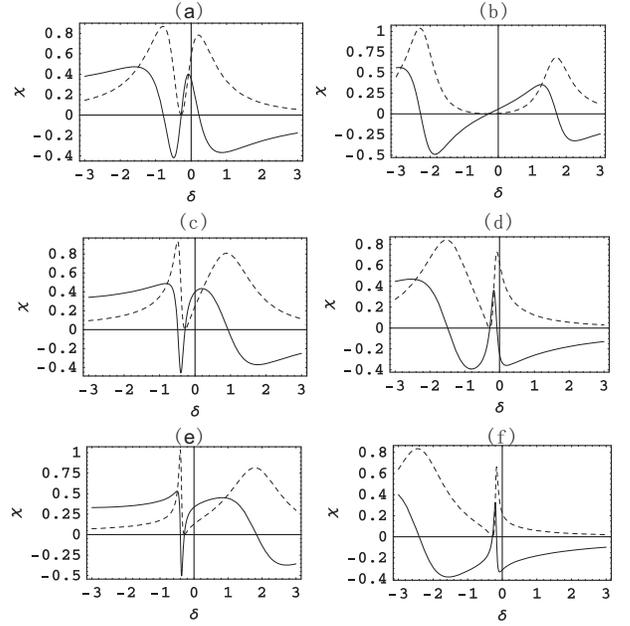}
\caption{Real (solid) and imaginary (dotted) parts of the linear
susceptibility as a function of normalized detuning $\protect\delta $ at $k=%
\protect\pi /4$. The parameters are set as $G_{1}=1$, $J=0.2$, $\ell =1$.
(a) $g_{2}=0.5$, $\protect\delta _{2}=0$; (b) $g_{2}=2$, $\protect\delta %
_{2}=0$; (c)$g_{2}=0.5$, $\protect\delta _{2}=1$; (d)$g_{2}=0.5$, $\protect%
\delta _{2}=-1$; (e)$g_{2}=0.5$, $\protect\delta _{2}=2$; (f)$g_{2}=0.5$, $%
\protect\delta _{2}=-2$. $\protect\delta $ is in units of $\protect\gamma%
_{A}=1$}
\label{fig:sus}
\end{figure}
%%%%%%%%%%%%%%%%%%%%%%%%%%%%%%%%%%%%%%%%%%%%%%%%%%%%%%%%%%%%%%%%%%%%%%%%%%%%%

In order to calculate the susceptibility, we first find the steady-state
solution by letting $\partial _{t}A_{k}=0$ and $\partial _{t}C_{k}=0$ in the
Eq.(\ref{eq}). The expectation value of $A_{k}$ over a stable state is
explicitly obtained as
\begin{equation}
\left\langle A_{k}\right\rangle =\frac{iG_{1}\left[ i\left( \omega
_{k}+\delta \right) -\gamma _{C}\right] }{\left[ i\left( \omega _{k}-\delta
_{1}\right) -\gamma _{A}\right] \left[ i\left( \omega _{k}+\delta \right)
-\gamma _{C}\right] +g_{2}^{2}}\left\langle a_{k}\right\rangle
\end{equation}%
where $\delta =\delta _{2}-\delta _{1}$. Since the coupling coefficient
\begin{equation*}
g_{1}=-\mu \sqrt{\frac{\omega _{0}}{2V\varepsilon _{0}}}\text{,}
\end{equation*}%
the real part $\chi _{k}^{r}$ and imaginary part $\chi _{k}^{i}$ of the
linear complex susceptibility $\chi _{k}$ are obtained as
\begin{eqnarray}
\chi _{k}^{r} &=&F\left[ \epsilon _{k}g_{2}^{2}-\left( \epsilon _{k}-\delta
_{1}\right) \left( \gamma _{C}^{2}+\epsilon _{k}^{2}\right) \right] L(k),
\notag \\
\chi _{k}^{i} &=&F\left[ \epsilon _{k}^{2}\gamma _{A}+\left( \gamma
_{A}\gamma _{C}+g_{2}^{2}\right) \gamma _{C}\right] L(k).
\end{eqnarray}%
where
\begin{eqnarray}
L(k)^{-1} &=&\left[ \gamma _{A}\gamma _{C}+g_{2}^{2}-\epsilon _{k}\left(
\epsilon _{k}-\delta _{2}\right) \right] ^{2}  \notag \\
&&+\left[ \epsilon _{k}\gamma _{A}+\left( \epsilon _{k}-\delta _{2}\right)
\gamma _{C}\right] ^{2}
\end{eqnarray}%
and $F=2G_{1}^{2}/\omega _{0}$.

Since the susceptibility depends on $k$, in Fig. \ref{fig:sus} the real and
imaginary susceptibilities $\chi _{k}^{r}$, $\chi _{k}^{i}$ are plotted
versus the detuning difference $\delta =\delta _{2}-\delta _{1}$ in units of
$\gamma _{A}$ ( $\gamma _{A}=10^{3}\gamma _{C}$), where we assume that the
central frequency of light pulse is at $k=\pi /4\ell $.

It is observed that, when the detuning $\delta _{1}$, $\delta _{2}$ satisfy $%
\epsilon _{k}=0$, that is, the two photon resonance is satisfied, both the
real and imaginary susceptibilities vanish, the absorption is absent and the
index of refraction is unity. Thus the whole system becomes transparent
under the driving of the strong classical control field. Through Eq. (\ref%
{eq:epsi}), we obtain that the momentum index $k$ together with the
nearest-neighbor evanescent coupling strength $J$ determines the range where
the transparency window occurs. The width of the transparency window depends
on the control field Rabi frequency $g_{2}$, which is shown by comparing
Fig. \ref{fig:sus} (a) with Fig. \ref{fig:sus} (b).

Finally to consider the role of the inter-cavity coupling $J$, we plot the
real (solid line) and the imaginary (dash line) part of the susceptibility
as a function of the inter-cavity coupling strength $J$, shown in Fig. \ref%
{fig:susJ}.
%%%%%%%%%%%%%%%%%%%%%%%%%%%%%%%%%%%%%%%%%%%%%%%%%%%%%%%%%%%%%%%%%%%%%%%%%%%%
\begin{figure}[tbp]
\includegraphics[width=8 cm]{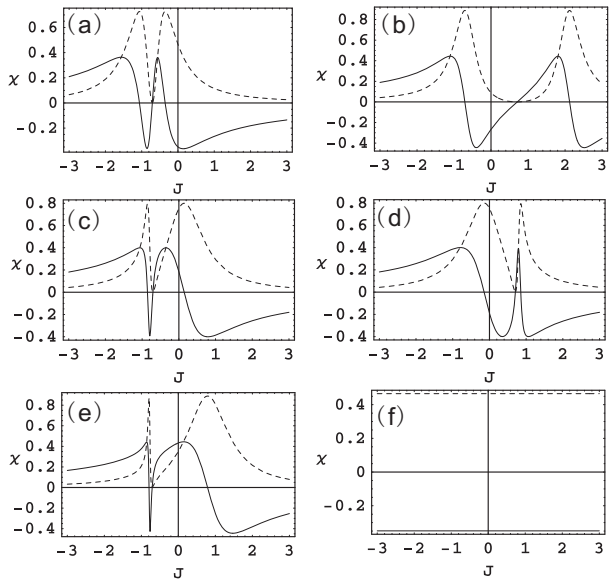}
\caption{Real (solid) and imaginary (dotted) parts of the linear
susceptibility as a function of the inter-cavity coupling strength $J$. The
parameters are set as $G_{1}=1$, $\ell =1$. (a) $g_{2}=0.5$, $\protect\delta %
_{2}=0$, $\protect\delta =1$ and $k=\protect\pi /4$; (b) $g_{2}=2$, $\protect%
\delta _{2}=0$, $\protect\delta =-1$ and $k=\protect\pi /4$; (c)$g_{2}=0.5$,
$\protect\delta _{2}=1$, $\protect\delta =1$ and $k=\protect\pi /4$; (d)$%
g_{2}=0.5$, $\protect\delta _{2}=-1$, $\protect\delta =-1$ and $k=\protect%
\pi /4$; (e)$g_{2}=0.5$, $\protect\delta _{2}=2$, $\protect\delta =1$ and $k=%
\protect\pi /4$; (f)$g_{2}=0.5$, $\protect\delta _{2}=0$, $\protect\delta =1$
and $k=\protect\pi /2$. $J$ is in units of $\protect\gamma _{A}=1$.}
\label{fig:susJ}
\end{figure}
%%%%%%%%%%%%%%%%%%%%%%%%%%%%%%%%%%%%%%%%%%%%%%%%%%%%%%%%%%%%%%%%%%%%%%%%%%%%%
It can be observed that: when the incident pulse is center at $k=\pi /(2\ell
)$, the susceptibility is independent of $J$ (see Fig. \ref{fig:susJ}(f));
for the input pulse centered at $k=\pi /(4\ell )$, in the vicinity of a
frequency corresponding to the two-photon Raman resonances, the medium made
of atoms becomes transparency with respect to the input pulse within the
photonic band. By comparing Fig. \ref{fig:susJ}(a) and (b), It can be found
that the detuning difference $\delta $ determines the position where the
transparency window occurs, and the intensity of the control beam decides
the width of the transparency window; it can also be observed from Fig. \ref%
{fig:sus} and Fig. \ref{fig:susJ} that the larger the detuning $|\delta _{2}|
$ is, the broader transparency window the spectra of this system has.

\section{\label{sec:five} Conclusion and remarks}

We have studied a hybrid system, which consists of $N$ homogeneously
coupled resonators with three-level $\Lambda $-type atoms doped in
each cavity. The electromagnetically induced transparency (EIT)
effect can enhance the ability for coherent manipulations on the
photon propagation in the CROW, namely, the photon transmission
along the CROW can be well controlled by the amplitude of the
driving field. With these results, it is expected that the quantum
information encoded in the input pulse can be stored and retrieved
by adiabatically tuning $g_{2}$ from $g_{2}\ll G_{1}$ to $g_{2}\gg
G_{1}$.

Also it can be seen that our hybrid architecture possesses more
controllable parameters for transferring photons in an array of
coupled cavities: two coupling strength $g_{1}$, $g_{2}$ and two
detunings. Typically, in two photon resonance, the light can be
stopped only by controlling the amplitude of the classical field. In
comparison with our scheme, the all optical architecture with the
passive optical resonator \cite{Fanprl04} only has two controllable
parameters, the coupling strength between the side-coupled cavity
and each of the CROW and the detuning between resonance frequency of
side-coupled cavity and that of cavity which constitute the CROW. In
other hand, the standard EIT approach only uses the single "cavity "
and thus there is not a controllable photonic band structure. The
on-chip periodic structure used here actually can implement the EIT
manipulation for
photonic storage in the periodic lattice fixing atoms spatially \cite%
{sunprl91}.

This work is supported by the NSFC with grant Nos. 90203018, 10474104 and
60433050, and NFRPC with Nos. 2006CB921206 and 2005CB724508. The author Lan
Zhou gratefully acknowledges the support of K. C. Wong Education Foundation,
Hong Kong.

\end{document}